\documentclass[preprint,10pt,5p,twocolumn]{elsarticle}

\usepackage{amssymb}
\usepackage{amsmath} 
\usepackage{url}

\usepackage{textcomp}

\journal{Computer Physics Communications}

\usepackage{color} 

\newcommand{\vb}[1]{\mathbf{#1}}

\begin{document}

\begin{frontmatter}

\title{Unravelling cosmic velocity flows: a Helmholtz-Hodge decomposition algorithm for cosmological simulations}

\author[inst1]{David Vall\'{e}s-P\'{e}rez} \corref{cor1} \ead{david.valles-perez@uv.es} 
\author[inst1]{Susana Planelles} \ead{susana.planelles@uv.es} 
\author[inst1,inst2]{Vicent Quilis} \ead{vicent.quilis@uv.es}
\cortext[cor1]{Corresponding author}

\address[inst1]{Departament d'Astronomia i Astrof\'{\i}sica, Universitat de
  Val\`encia, E-46100 Burjassot (Val\`encia), Spain}
\address[inst2]{Observatori Astron\`omic, Universitat de Val\`encia, E-46980
  Paterna (Val\`encia), Spain}

\begin{abstract}
In the context of intra-cluster medium turbulence, it is essential to be able to split the turbulent velocity field in a compressive and a solenoidal component. We describe and implement a new method for this aim, i.e., performing a Helmholtz-Hodge decomposition, in multi-grid, multi-resolution descriptions, focusing on (but not being restricted to) the outputs of AMR cosmological simulations. The method is based on solving elliptic equations for a scalar and a vector potential, from which the compressive and the solenoidal velocity fields, respectively, are derived through differentiation. These equations are addressed using a combination of Fourier (for the base grid) and iterative (for the refinement grids) methods. We present several idealised tests for our implementation, reporting typical median errors in the order of 1\textperthousand-$1\%$, and with 95-percentile errors below a few percents. Additionally, we also apply the code to the outcomes of a cosmological simulation, achieving similar accuracy at all resolutions, even in the case of highly non-linear velocity fields. We finally take a closer look to the decomposition of the velocity field around a massive galaxy cluster.
\end{abstract}



\end{frontmatter}


\section{Introduction}
\label{s:intro}

Cosmological structures and, in particular, galaxy clusters, which constitute the most massive structures which have had time to collapse under their own gravity, are dynamically interesting objects from several perspectives. The non-linearity of the evolution of their baryonic component (i.e., ordinary matter, most of which is in the form of a hot, tenuous plasma known as the intra-cluster medium, ICM) couples the different scales in the evolution of cosmic inhomogeneities, producing a plethora of complex hydrodynamical phenomena, such as shock waves and turbulence. Turbulence is an intrinsically multi-scale phenomenon, since bulk motions trigger (magneto-)hydrodynamical instabilities (e.g., \cite{ZuHone_2011}) which cascade down to smaller scales until they get dissipated by viscous effects. 

Turbulent motions can be \cite{Gaspari_2013} and have been recently \cite{Zhuravleva_2014, Hofmann_2016} measured on a number of nearby galaxy clusters from X-ray surface brightness fluctuations, which have in turn been connected to signatures of particle acceleration and diffuse radio emission in the ICM \cite{Eckert_2017, Bonafede_2018}. Future X-ray facilities, like \textsc{Athena}\footnote{\url{www.the-athena-x-ray-observatory.eu}}, will offer unprecedented insight into the dynamics of turbulent motions in the ICM \cite{Roncarelli_2018}. However, the precise theoretical and numerical description of turbulence in these vast structures is still matter of ongoing research (e.g., \cite{Miniati_2015, Schmidt_2016, Vazza_2017, Iapichino_2017}, just to highlight a few recent numerical studies).

Splitting the (turbulent) velocity field in its compressive and solenoidal components, i.e., performing a Helmholtz-Hodge decomposition (HHD), is a crucial step towards exploring the role of turbulence in the ICM, since these components play fundamental and distinct roles in the evolution of cosmic structures. Thus, while the solenoidal component is likely the major responsible for the amplification of cosmic magnetic fields \cite{Vazza_2014, Porter_2015}, compressive turbulence has an important role in generating weak shocks which have consequential effects on the magnetic and thermal evolution of the cluster (e.g., \cite{Porter_2015, Federrath_2011}). These two components also differ in their spatial distribution, the former tending to be more volume-filling \cite{Vazza_2017, Federrath_2009, Iapichino_2011}, and even in their spectrum, steeper for the compressive component \cite{Miniati_2015, Federrath_2010, Federrath_2013, Miniati_2014}. The distinction between the solenoidal and compressive component of the velocity field is also of utmost importance to model the acceleration of cosmic ray particles in the ICM, which will become a vibrant field of observational research with the advent of the new generation of radio telescopes (e.g., SKA \cite{Acosta-Pulido_2015}). In this line, a lot of effort has been recently put in modelling the acceleration by compressive and solenoidal modes in magnetohydrodynamics (MHD; \cite{Fujita_2003, Cassano_2005, Brunetti_2011, Pinzke_2017, Brunetti_2020}).

Cosmological hydrodynamical simulations, as many other applications in computer fluid dynamics, need a huge dynamical range to be resolved in order to, for example, form realistic galaxies in a cosmological environment \cite{Bertschinger_1998, Naab_2017} or adequately describe turbulence in the ICM (see, e.g., \cite{Vazza_2012} for some graphical examples of the effects of resolution on the ability to capture instabilities; see also \cite{Mohapatra_2020, Mohapatra_2021}, who present detailed studies of stratified, ICM-like turbulence in numerical grids of varying resolution). While Lagrangian codes are inherently adaptive, Eulerian codes based on high-resolution shock-capturing (HRSC) techniques are especially capable of handling shocks and other types of discontinuities, which are pervasive in the formation of cosmological structures (\cite{Bykov_2008}, for a review). That is why, among several other options (see, for example, \cite{Dolag_2008} for a broad review), Adaptive Mesh Refinement (AMR) codes are especially suited for cosmological structure formation.

Previous studies of ICM turbulence have already implemented HHD algorithms. For example, several works using uniform grids (or fixed refinement strategies which, ultimately, allow to resolve the object of interest within a constant resolution) perform the decomposition in Fourier space, where it simply reduces to linear algebra projections (e.g., \cite{Vazza_2017, Kritsuk_2011}). Additionally, \cite{Vazza_2017} also confront this method with solving a Poisson equation (in Fourier space) to find a scalar potential for the compressive velocity component, reporting more accurate results for the first method. However, any of these two procedures, because of their usage of FFT algorithms, can only be applied to regular, uniform grids, which are not the common use in cosmological simulations.

In this paper, we propose, implement and test a new algorithm for performing the Helmholtz-Hodge decomposition in a multi-scale AMR grid, which can therefore be applied to the outcomes of a full-cosmological simulation without the need of performing resimulations of specific objects or constrained simulations. Our method decomposes the velocity fields by solving elliptic partial derivative equations (PDEs), which can be addressed iteratively using a wide variety of well-known algorithms (e.g., \cite{Press_1992}). Nevertheless, although our primary focus is the application to cosmological structure formation, we emphasize that the approach presented in this work can be directly applied to any application of block-structured AMR, and readily extended to particle-based simulations through a suitable interpolation scheme.

The rest of the manuscript is organised as follows. In Sec. \ref{s:method}, we describe our method for performing the decomposition and discuss its numerical implementation. In Sec. \ref{s:tests} we present and describe a set of tests to validate the accuracy of the algorithm, while in Sec. \ref{s:tests_masclet} we apply it to the complex velocity field of a cosmological simulation. Last, in Sec. \ref{s:conclusion} we summarise and present our conclusions.

\section{Description of the method}
\label{s:method}

The algorithm is based on the Helmholtz-Hodge decomposition (see, e.g., \cite{Arfken_2013, Kowal_2010, Harouna_2012}), which allows to univocally split any velocity field in three terms,

\begin{equation}
	\vb{v} = \vb{v}_\mathrm{comp} + \vb{v}_\mathrm{rot} + \vb{v}_\mathrm{harm}
	\label{eq:velocity_3terms}
\end{equation}

\noindent where $\vb{v}_\mathrm{comp}$ is the compressive (or irrotational, $\nabla \times \vb{v}_\mathrm{comp} = \vb{0}$) velocity field, $\vb{v}_\mathrm{rot}$ is the purely rotational (or solenoidal, $\nabla \cdot \vb{v}_\mathrm{rot} = 0$) velocity field and $\vb{v}_\mathrm{harm}$ is the harmonic velocity field (both irrotational and solenoidal, thus satisfying $\nabla^2 \vb{v}_\mathrm{harm} = \vb{0}$ and $\vb{v}_\mathrm{harm} = \nabla \chi$, with $\nabla^2\chi = 0$). 

The harmonic component can be shown to be identically null in any domain with periodic boundary conditions, and therefore we will no longer consider it. Because of their defining properties, the compressive component can be written as the gradient of a scalar potential, $\vb{v}_\mathrm{comp} = - \nabla \phi$, while the rotational component can be derived from a vector potential, $\vb{v}_\mathrm{rot} = \nabla \times \vb{A}$. From this, it is easy to derive that the scalar and vector potentials can be computed, respectively, as the solutions of the following elliptic PDEs,

\begin{equation}
	\nabla^2 \phi = - \nabla \cdot \vb{v}
	\label{eq:poisson_scalar}
\end{equation}
\begin{equation}
	\nabla^2 \vb{A} = - \nabla \times \vb{v}
	\label{eq:poisson_vector}
\end{equation}

\noindent which are formally equivalent to a set of four decoupled Poisson equations (one for $\phi$ and one for each cartesian component of $\vb{A}$) whose sources are the divergence and the components of the curl of the overall velocity field. 

Once the potentials have been obtained, the compressive and rotational components of the velocity field can be obtained through differentiation. Note that, in principle, it would only be necessary to find one of the potentials (either $\phi$ or $\vb{A}$), since the other velocity component could be then derived by subtracting to the total velocity (Eq. \ref{eq:velocity_3terms} with $\vb{v}_\mathrm{harm} = \vb{0}$). Nevertheless, we have chosen to compute all the potentials in order to keep track of the associated numerical errors.

\subsection{Numerical implementation}
\label{s:methods.numerics}
We have designed a code to perform such decomposition in a multi-resolution, block-structured AMR velocity field. As mentioned before, our code can be easily applied to the outcomes of any AMR simulation (cosmological or not), or even to a particle-based code, from which the continuous velocity field can be defined on an ad-hoc AMR grid structure through a smoothing method (e.g., a particle-mesh, \cite{Hockney_1981}).

In our particular implementation, for the base (coarsest) level, $\ell = 0$, taking advantage of the periodic boundary conditions, Poisson equations are solved in Fourier space\footnote{Note, however, that this is not a requirement of the method. We solve Poisson's equations in Fourier space as cosmological simulations of sufficiently large volumes typically implement periodic boundary conditions and, in these situations, solving Poisson's equation in Fourier space is much more computationally efficient than using iterative methods. In any case, for non-periodic domains, the base level can be addressed through iterative schemes, just as described below for the refinement levels, if suitable boundary conditions are provided. In the case of non-periodic domains, however, the harmonic term cannot be dropped in Eq. \ref{eq:velocity_3terms}. While we cannot compute $\vb{v}_\mathrm{harm}$ by solving elliptic equations, this term can be obtained just by using Eq. \ref{eq:velocity_3terms} to solve for it, once $\vb{v}_\mathrm{comp}$ and $\vb{v}_\mathrm{rot}$ have been found. This is a good approach, as long as our algorithm precisely reconstructs the compressive and rotational velocity fields from an input field, which is tested in Sec. \ref{s:tests} and Sec. \ref{s:tests_masclet}.}, where they just reduce to a multiplication by the Green's function. The basic procedure followed to solve Poisson equations in the base grid can be described in the following steps:

\begin{enumerate}
	\item The fast Fourier transform (FFT) of the source (right-hand side terms in Eqs. \ref{eq:poisson_scalar} or \ref{eq:poisson_vector}) in the base grid is computed, yielding a set of coefficients $F_{lmn}$.
	\item Poisson's equation is then solved in Fourier space by multiplying by the Green's function, $G_{lmn}$: $\, \tilde \phi_{lmn} = G_{lmn} F_{lmn}$. The Green's function is given by:
	\begin{equation}
		G_\mathrm{lmn} = \frac{(\Delta x / 2)^2}{\sin^2 \left(\frac{\pi l}{N_x}\right) + \sin^2 \left(\frac{\pi m}{N_y}\right) + \sin^2 \left(\frac{\pi n}{N_z}\right)}
		\label{eq:green_function_poisson}
	\end{equation}
	\noindent where $\Delta x$ is the cell side length and the domain is discretised in {$N_x \times N_y \times N_z$} cells \cite{Hockney_1981}.
	\item The inverse FFT of $\tilde \phi_{lmn}$ yields the sought potential at the base level.
\end{enumerate}

In subsequent refinement levels, $\ell > 0$, Poisson equations have to be solved taking into account the boundary conditions imposed by the coarser grids the refined patches are embedded into. In order to do so, we use a successive over-relaxation procedure (SOR; see, for example, \cite{Press_1992}) on the discretised Poisson equation. 

Each AMR patch is first initialised (both in the boundary and in the interior cells) by linear interpolation from the values of the potential at the best-resolved lower level patch available. Then, the interior cells are iteratively updated in a chessboard pattern as

\begin{equation}
	\phi_{i, j, k}^\mathrm{new}=\omega \phi_{i, j, k}^{*}+(1-\omega) \phi_{i, j, k}^\mathrm{old}
	\label{eq:sor_1},
\end{equation}

\noindent where

\begin{equation}
	\begin{array}{c}
		\phi_{i, j, k}^{*}=\frac{1}{6}\left[\phi_{i+1, j, k}^\mathrm{old}+\phi_{i-1, j, k}^\mathrm{old}+\phi_{i, j+1, k}^\mathrm{old}+\phi_{i, j-1, k}^\mathrm{old}\right. \\
		\left.+\phi_{i, j, k+1}^\mathrm{old}+\phi_{i, j, k-1}^\mathrm{old} - (\Delta x_\ell)^2 f_{i,j,k}\right],
	\end{array}
	\label{eq:sor_2}
\end{equation}

\noindent being $\Delta x_\ell$ the cell size at the given refinement level, $f_{i,j,k}$ the source term and $1<\omega<2$ the over-relaxation parameter. In order to boost convergence, we set $\omega$ according to the Chebyshev acceleration procedure \citep{Press_1992}. Aiming to avoid undesirable boundary effects due to the interpolation of the potential boundary conditions, we extend the patches with 3 fictitious cells in all directions, so that these boundary conditions are enforced slightly far away from the region of interest.

Once $\phi$ and $\vb{A}$ are known, the velocity components are found by finite differencing the potentials as defined before. We compute the derivatives (both of the velocity and of the potentials) using an eighth-order scheme with a centered stencil of up to\footnote{The stencil length is shortened as cells get closer to the boundary.} 9 points, which provides robust values of the differential operators of the velocity fields and mitigates the impact of spurious noise. 

The code is parallelised according to the OpenMP standard directives. Our implementation is freely available through its GitHub repository\footnote{\url{https://www.github.com/dvallesp/vortex/}}.

\section{Tests}
\label{s:tests}

Aiming to assess the robustness of our HHD method and its implementation, we have designed a battery of tests focused on quantifying to which extent the code is able to accurately identify and disentangle the compressive and rotational velocities. We describe such tests in Sec. \ref{s:tests.description}, and examine their results in Sec. \ref{s:tests.results}.

\subsection{Description of the tests}
\label{s:tests.description}

For the four tests described below, we have first established a simple AMR grid structure, i.e., a set of patches at different refinement levels that could reasonably mock the ones generated in an actual simulation.

We have considered a cubic domain of unit length, with origin at the center of the box, and we have discretised it with $128^3$ cells. For each octant, we establish a first refinement patch with twice the spatial resolution covering the central $1/8$ of the octant volume. For example, in the first octant ($0<x,y,z<1/2$), we set up a patch at $1/8 < x,y,z < 3/8$ with $64^3$ cells (and likewise for the remaining 7 octants). Then we add a $\ell = 2$ patch in the central $1/8$ volume of each $\ell = 1$ patch, and continue recursively.

\begin{figure}
	\centering
	\includegraphics[width=\linewidth]{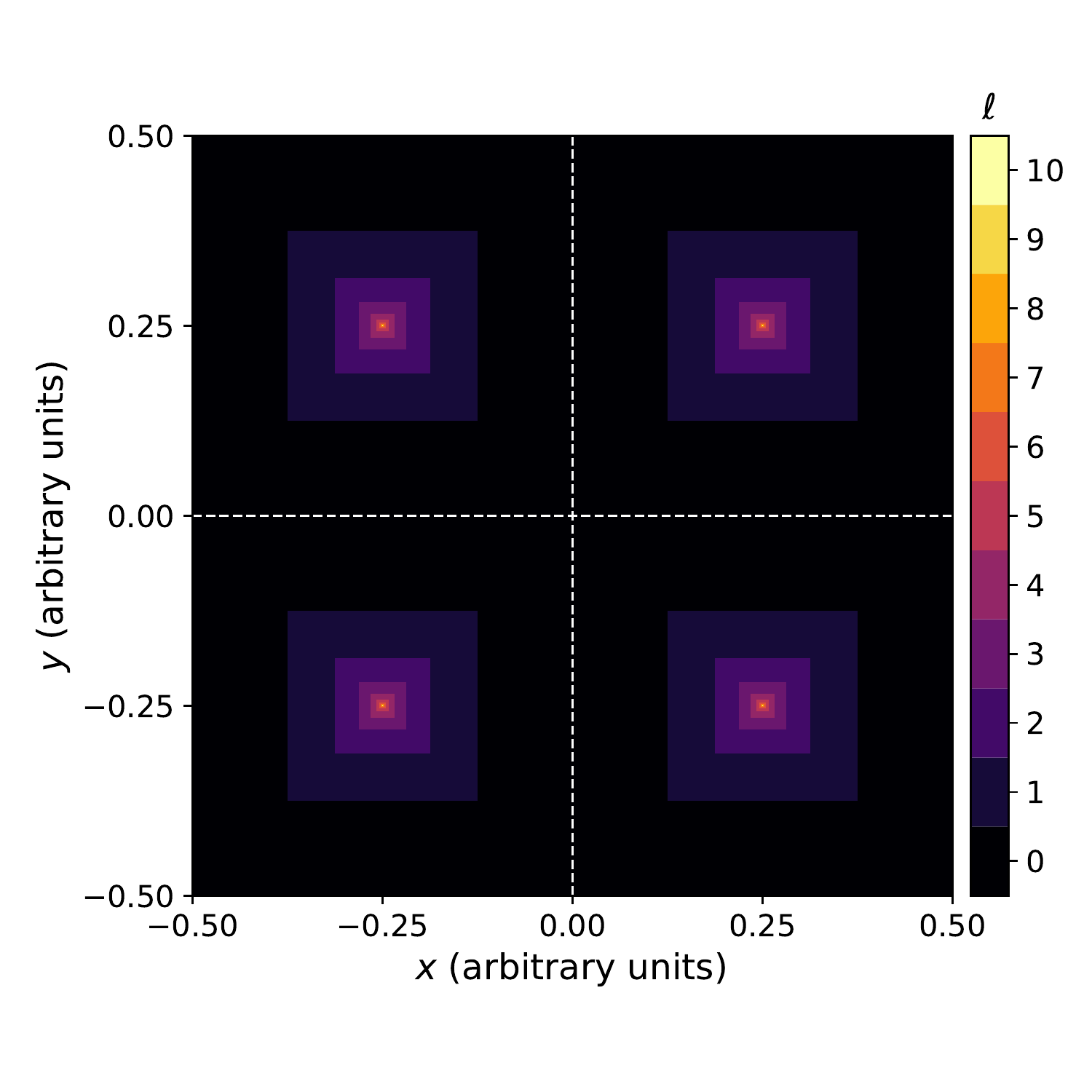}
	\caption{AMR grid structure for the tests in Sec. \ref{s:tests}. The figure represents a slice at $z=0.25$, with the colors encoding the highest refinement level at each point in the $x$--$y$ plane. The white, dashed lines indicate the cartesian $x$ and $y$ axes.}
	\label{fig:test_grid}
\end{figure}

In the first three tests we consider up to $\ell_\mathrm{max} = 10$ refinement levels, providing a peak resolution of $7.63 \times 10^{-6}$ (relative to the box size, which is normalised to $1$). Figure \ref{fig:test_grid} presents graphically the grid structure employed in these tests. Below, we describe the velocity fields that we have seeded on these grids.

\subsubsection{Test 1: constant divergence field}
\label{s:tests.description.1}
For the first test, we consider a purely compressive velocity field with constant divergence, given in cartesian and spherical coordinates by the analytical expression

\begin{equation}
	\vb{v} = \omega_0 \left(x \,\vb{\hat u_x} + y \,\vb{\hat u_y} + z \,\vb{\hat u_z} \right) \equiv \omega_0 r \, \vb{\hat u_r}.
	\label{eq:v_test1}
\end{equation}

This field has $\nabla \cdot \vb{v} = 3 \omega_0$ and $\nabla \times \vb{v} = \vb{0}$, and hence $\vb{v} = \vb{v}_\mathrm{comp}$ and $\vb{v}_\mathrm{rot} = 0$. It is easy to find an analytic expression for the scalar potential, $\phi = - \frac{\omega_0}{2} \left(x^2 + y^2 + z^2\right) + \mathcal{C}$, with $\mathcal{C}$ any arbitrary real constant. Likewise, the vector potential ought to be $\vb{A} = \nabla \chi$, with $\chi$ an arbitrary scalar function or, in particular, $\vb{A} = \vb{0}$. We have set $\omega_0 = 0.01$.

\subsubsection{Test 2: constant curl field}
\label{s:tests.description.2}
Analogous to the previous one, we have also considered the case of a purely uniformly rotating velocity field, which can be analytically given in cartesian and cylindrical coordinates as:

\begin{equation}
	\vb{v} = \omega_0 \left(- y \, \vb{\hat u_x} + x \, \vb{\hat u_y} \right) \equiv \omega_0 \rho \, \vb{\hat u_\phi}
	\label{eq:v_test2}
\end{equation}

It is straightforward to show that $\nabla \cdot \vb{v} = 0$ and $\nabla \times \vb{v} = 2 \omega_0 \, \vb{\hat u_z}$. This velocity field is generated by the potentials $\phi = \mathcal{C}$, being $\mathcal{C} \in \mathbb{R}$ a free constant, and $\vb{A} = - 2 \omega_0 \rho \, \vb{\hat u_z} + \nabla \chi$, with $\chi$ any arbitrary scalar function. As in Test 1, we set $\omega_0 = 0.01$.

These two previous tests (Test 1 and Test 2) are aimed to validate the reconstruction of pure velocity fields (either compressive or rotational, respectively), i.e., to estimate the magnitude of the errors involved in the procedure described in Sec. \ref{s:methods.numerics} in idealised situations where there is no cross-talk between rotational and compressive components. Last, note that even though these two velocity fields are not periodic in the mock simulation domain, this will only affect a negligible amount of cells in the domain's boundary.

\subsubsection{Test 3: compressive and rotational, periodic field}
\label{s:tests.description.3}
We have designed a third test, aimed to assess the effects of the cross-talk between the compressive and the rotational components. We have considered the velocity field:

\begin{equation}
	\begin{aligned}
		\vb{v} =&\left[\sin\left(2\pi x\right) + \sin\left(4\pi y\right) + \sin\left(6\pi z\right)\right] \,\vb{\hat{u}_x} \;+ \\
		+&\left[\sin\left(6\pi x\right) + \sin\left(2\pi y\right) + \sin\left(4\pi z\right)\right] \,\vb{\hat{u}_y} \;+ \\
		+&\left[\sin\left(4\pi x\right) + \sin\left(6\pi y\right) + \sin\left(2\pi z\right)\right] \,\vb{\hat{u}_z}.
		\label{eq:v_test3}
	\end{aligned}
\end{equation}

The compressive part corresponds to the terms of angular frequency $2\pi$, while the higher frequency ones constitute the rotational component. Also in this case, it is easy to find analytical solutions to $\phi$ and $\vb{A}$:

\begin{equation}
	\phi = \frac{1}{2\pi} \left[ \cos\left(2\pi x\right) + \cos\left(2\pi y\right) + \cos\left(2\pi z\right) \right] + \mathcal{C}
	\label{eq:phi_test3}
\end{equation}

\begin{equation}
	\begin{aligned}
		\vb{A} =&\left[- \frac{1}{4\pi} \cos\left(4\pi z \right) + \frac{1}{6\pi} \cos\left(6\pi y \right)\right] \,\vb{\hat{u}_x} \;+ \\
		+&\left[- \frac{1}{4\pi} \cos\left(4\pi x \right) + \frac{1}{6\pi} \cos\left(6\pi z \right) \right] \,\vb{\hat{u}_y} \;+ \\
		+&\left[- \frac{1}{4\pi} \cos\left(4\pi y \right) + \frac{1}{6\pi} \cos\left(6\pi x \right)\right] \,\vb{\hat{u}_z} + \nabla \chi.
		\label{eq:A_test3}
	\end{aligned}
\end{equation}

\subsubsection{Test 4: ICM-like mock velocity field}
\label{s:tests.description.4}

While the previous tests have checked the ability of the code to reconstruct idealised solenoidal, compressive and mixed velocity fields, we have implemented a last test aimed to assess the ability of the code to capture and reconstruct variations on a broad range of spatial frequencies. The test is, in part, inspired by the one presented by \cite[App. A.1.2]{Vazza_2017}, but with some differences aimed to mix both velocity fields (instead of generating a purely solenoidal or compressive field), while still having an analytic solution of the HHD to compare with the numerical results.

We have generated our mock, ICM-like velocity field according to the procedure described below: 
	
	\begin{enumerate}
		\item We consider a uniform grid of $(N_x \times 2^{\ell_\mathrm{max}})^3$ cells. In that grid, we compute the velocity field $\vb{v} = \vb{v}_\mathrm{comp} + \vb{v}_\mathrm{rot}$, with:
		\begin{equation}
			\vb{v}_\mathrm{comp} = \sum_{i=x,y,z} \sum_{n=N_\mathrm{min}}^{N_\mathrm{max}} A^\mathrm{comp}_n \sin \left( \frac{2\pi n}{L} x_i  + \psi_\mathrm{n}^\mathrm{comp,i}\right) \vb{\hat u_i}
			\label{eq:test4_vcomp}
		\end{equation}
		\begin{equation}
			\vb{v}_\mathrm{rot} = \sum_{i=x,y,z} \sum_{j \neq i} \sum_{n=N_\mathrm{min}}^{N_\mathrm{max}} A^\mathrm{rot}_n \sin \left( \frac{2\pi n}{L} x_j  + \psi_\mathrm{n}^\mathrm{rot,ij}\right) \vb{\hat u_i}
			\label{eq:test4_vrot}
		\end{equation}
		\noindent being $L$ the box side length ($L=1$ in our case), $A_n^\mathrm{comp}$ ($A_n^\mathrm{rot}$) the amplitude of the mode of frequency $n$ of the compressive (solenoidal) component, and $\psi_\mathrm{n}^\mathrm{comp,i}$ ($\psi_\mathrm{n}^\mathrm{rot,ij}$) the initial phases. For the intial phases, we have generated 9 sets of $N_\mathrm{max} - N_\mathrm{min} + 1$ random numbers, uniformly sampled from the interval $0 < \psi < 2 \pi$. The amplitudes are generated so that the compressive (solenoidal) component follows a Burgers \cite{Burgers_1939} (Kolmogorov \cite{Kolmogorov_1941}) spectrum, $E(k) \propto k^{-2}$ ($E(k) \propto k^{-5/3}$). We set $N_\mathrm{min} = 2$ and $N_\mathrm{max} = 1024$, so that our mock velocity field presents solenoidal and compressive fluctuations over scales differing almost 3 orders of magnitudes. In order for both components to be relevant, we fix the amplitudes so that $A_n^\mathrm{rot}=A_n^{\mathrm{comp}}$ for $n=64$. We note that, while the velocity field generated according to this procedures is not the most general one (e.g., one could add oblique plane waves), it is challenging enough in order to show the capability of our code to handle a broad range of spatial frequencies in close-to-realistic conditions.
		\item Then, we compute this total velocity field onto the AMR grid structure defined at the beginning of Sec. \ref{s:tests.description}. In this case, we maintain $N_x = 128$ as the resolution of the base grid, and limit the number of refinement patches to\footnote{The limitation is due to the fact that we need to compute the velocity field in a uniform grid, in the first place. Thus, with $N_x = 128$ and $\ell_\mathrm{max} = 4$, this uniform grid consists of $2048^3$ cells.} $\ell_\mathrm{max} = 4$, which is still enough to show the multi-scale capabilities of our code. When computing the value of the velocities on the base level or on AMR levels $\ell < \ell_\mathrm{max}$, we average over the uniform grid cells enclosed in the coarser volume element.
	\end{enumerate}
	
This process consistently generates a mixed, solenoidal and compressive, velocity field presenting similar scaling features as the ICM over almost 3 decades in spatial frequency. Therefore, it can robustly show the capability of the code to handle multi-scale (and multi-frequency) velocity signals.

\subsection{Results}
\label{s:tests.results}
For each test, we have validated the performance of the code by computing a series of error statistics, which we present below. Let $\vb{v}$ be the input velocity field, for which our algorithm returns its compressive and rotational components, $\vb{\tilde v}_\mathrm{comp}$ and $\vb{\tilde v}_\mathrm{rot}$. Thus, the algorithm recovers a total velocity field $\vb{\tilde v} = \vb{\tilde v}_\mathrm{comp} + \vb{\tilde v}_\mathrm{rot}$ which might differ from the original, $\vb{v}$, due to the numerical error in the processes involved, namely finite-differencing the velocity field, integrating the elliptical equations and finite-differencing the potentials.

In Tests 1 and 2, where only one velocity component (compressive and rotational, respectively) was present, we have quantified the error in reconstructing these velocity fields by computing the cell-wise relative error as\footnote{The equation below corresponds to the propagation of the variance in $\{v_i^\mathrm{cell}\}$ to the function $v^\mathrm{cell} = \sqrt{\sum_{i=1}^3 \left(v_i^\mathrm{cell}\right)^2}$, assuming the velocity components are uncorrelated.}:

\begin{equation}
	\varepsilon_r \left(v_\mathrm{cell} \right) = \sqrt{\sum_{i=1}^3 \left[\left( \frac{v_\mathrm{cell}^i}{|\vb{v}_\mathrm{cell}|} \right)^2 \left|\frac{\tilde{v}^i_\mathrm{cell} - v^i_\mathrm{cell}}{v^i_\mathrm{cell} + \epsilon}\right|\right]^2}
	\label{eq:relative_error}
\end{equation}

\noindent where the subindex $i$ runs over the three cartesian components. Note that we add a small constant, $\epsilon$, in the denominator of the relative error in $v_i^\mathrm{cell}$ to prevent the overestimation of the error due to the cells with velocities close to 0. We set $\epsilon \equiv 10^{-2} \max \left(|v_i|\right)$. For each refinement level, $\ell$, we have computed the median error over all the cells, which we use as an estimate of the velocity reconstruction error. We also consider the error percentiles 5, 25, 75 and 95 in order to give a confidence interval (CI) for this error.

The velocity field in Test 1 (Test 2) has null rotational (compressive) component. We have checked this by computing the median value and the 5, 25, 75 and 95 percentiles of $|\vb{v}_\mathrm{rot}|/|\vb{v}|$ ($|\vb{v}_\mathrm{comp}|/|\vb{v}|$).

\begin{figure}
\centering
\includegraphics[width=\linewidth]{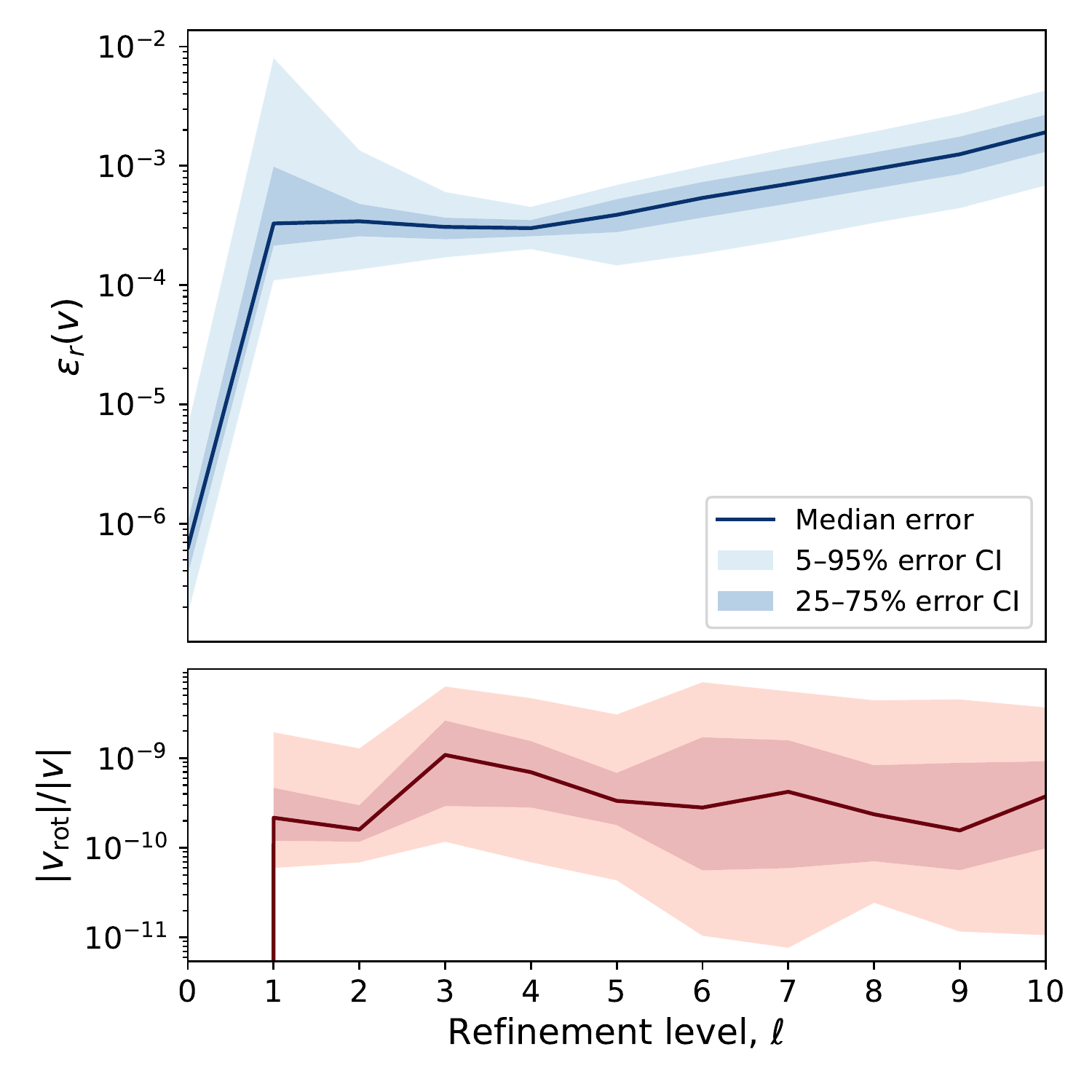}
\caption{Results from Test 1. \textit{Upper panel}: median relative error (solid line; defined as in Eq. \ref{eq:relative_error}) and confidence intervals (CIs) in the reconstruction of the velocity field. \textit{Lower panel}: fraction of rotational velocity misreconstructed by the algorithm.}
\label{fig:test1_results}
\end{figure}

\begin{figure}
\centering
\includegraphics[width=\linewidth]{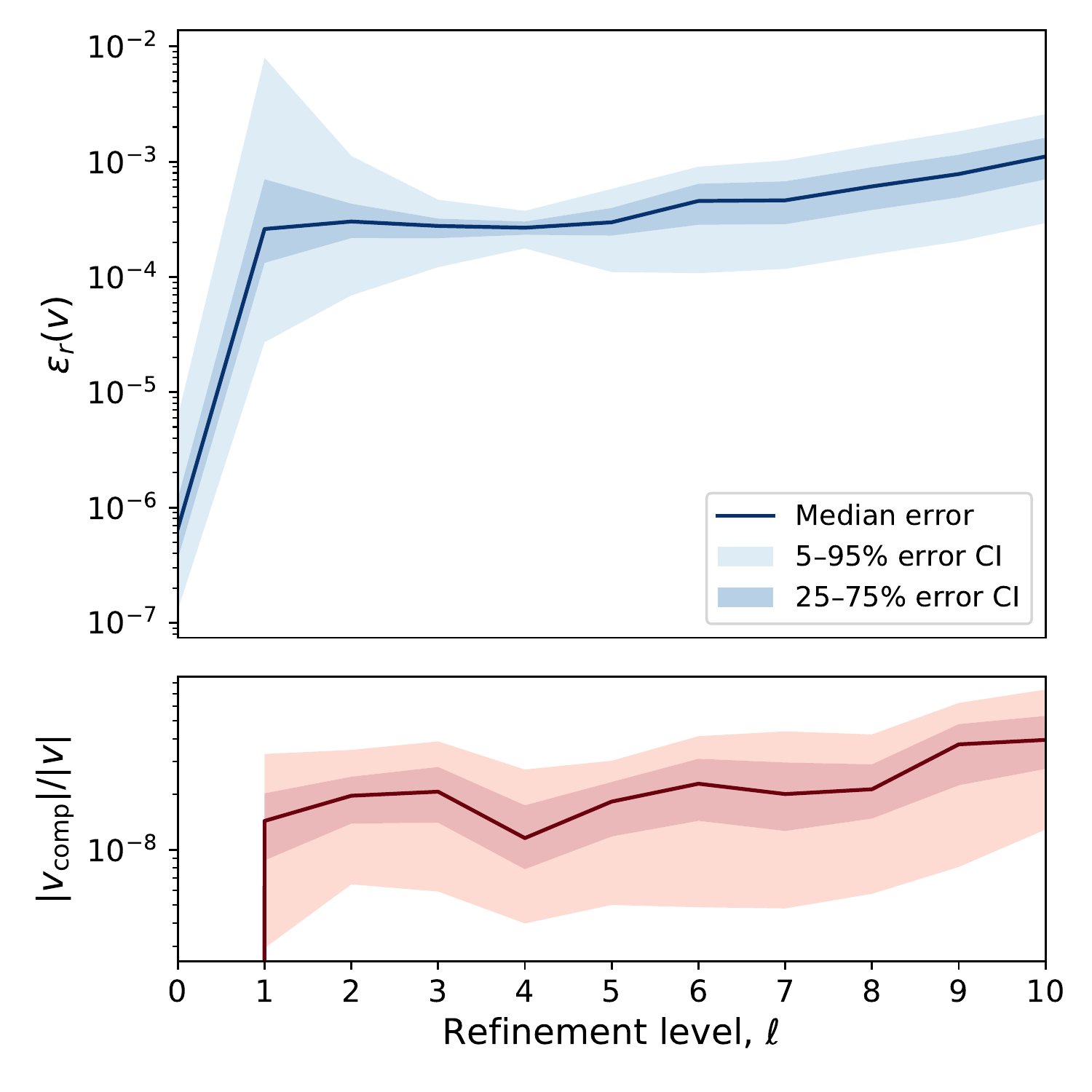}
\caption{Results from Test 2. \textit{Upper panel}: median relative error (solid line; defined as in Eq. \ref{eq:relative_error}) and confidence intervals (CIs) in the reconstruction of the velocity field. \textit{Lower panel}: fraction of compressive velocity misreconstructed by the algorithm.}
\label{fig:test2_results}
\end{figure}

The results for Test 1 and Test 2 are summarised in Figures \ref{fig:test1_results} and \ref{fig:test2_results}, respectively. The upper panels show $\varepsilon_r(v)$ as defined above. Both in the constant divergence and the constant rotational test, the median errors are small (typically lower than $1$\textperthousand), and even the 95-percentile errors do not exceed $1\%$ at any refinement level. The base level, for which FFT is used, has much more precise results, with median relative errors below $10^{-6}$. For the AMR levels, we only find a very slight trend to increase the error in more refined levels.

The lower panels in Figs. \ref{fig:test1_results} and \ref{fig:test2_results} show $|\vb{v}_\mathrm{rot}|/|\vb{v}|$, for Test 1, and $|\vb{v}_\mathrm{comp}|/|\vb{v}|$, for Test 2. For these highly-idealised scenarios, there is virtually no cross-talk between the different components: the rotational velocity in Test 1 accounts for less than $10^{-8}$ in more than $95\%$ of the cells at any level. Likewise, more than $95\%$ of the cells in Test 2 have compressive velocities less than $10^{-7}$ relative to the total velocity magnitude.

\begin{figure}
\centering
\includegraphics[width=\linewidth]{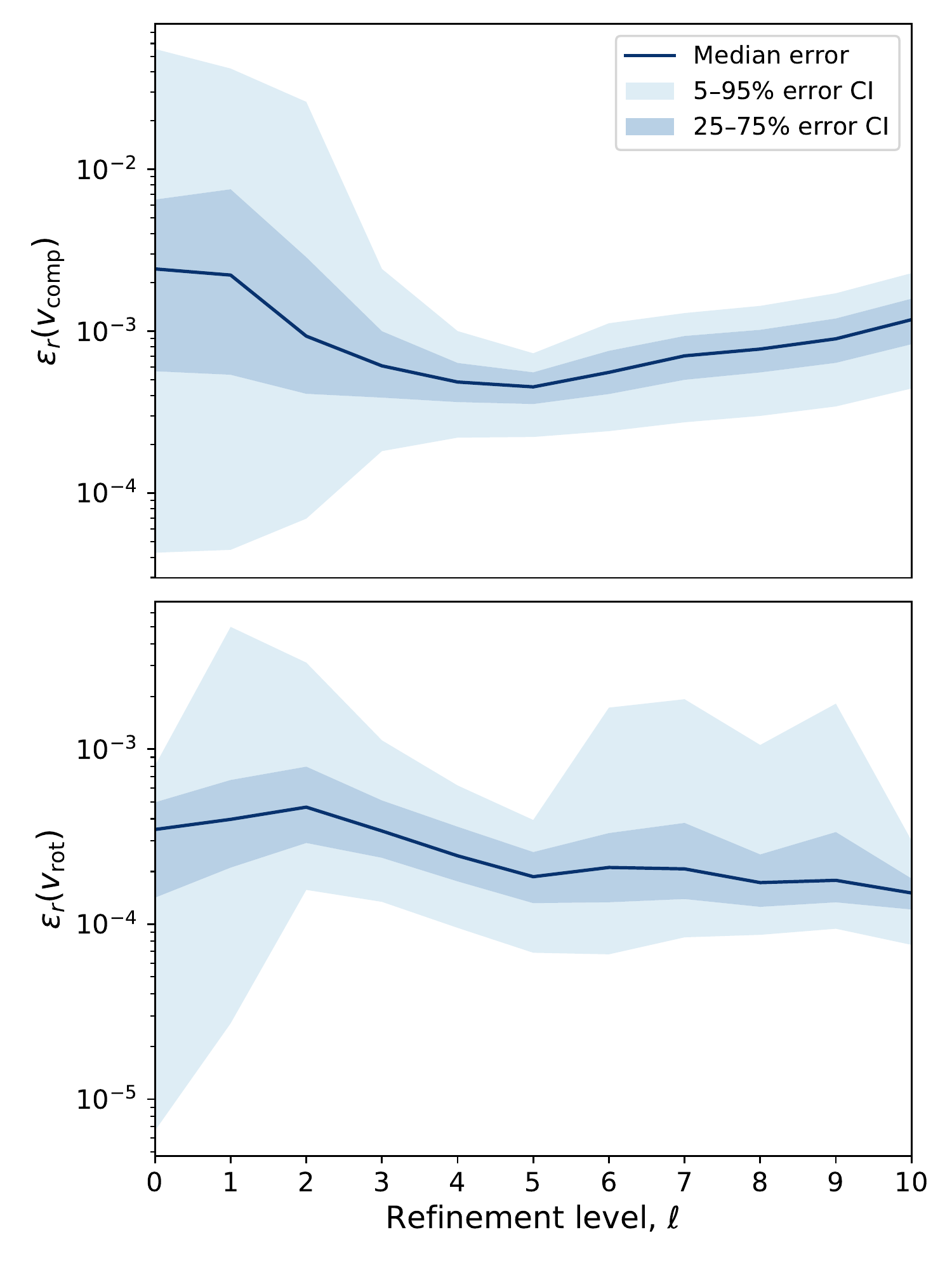}
\caption{Results from Test 3. Upper (\textit{lower}) panel presents the relative error in the reconstruction of the compressive (\textit{rotational}) velocity fields.}
\label{fig:test3_results}
\end{figure}

In Test 3, as both the compressive and the rotational velocity components are present, we have quantified the relative error in disentangling and reconstructing each of these by applying Eq. (\ref{eq:relative_error}) separately to each component. We present its results in Figure \ref{fig:test3_results}.

For the compressive velocity, the typical relative errors are below $3 \times 10^{-3}$, while the rotational velocity presents median relative errors in the range $10^{-4}$--$10^{-3}$. For the AMR levels, $\ell \geq 1$, the magnitude and behaviour of the errors resembles what has been seen before for Tests 1 and 2. Remarkably, in this example where both components are present, the base level does no longer exhibit a much more precise result, but it shows errors in the same order as the ones in the AMR levels.

\begin{figure}
\centering
\includegraphics[width=\linewidth]{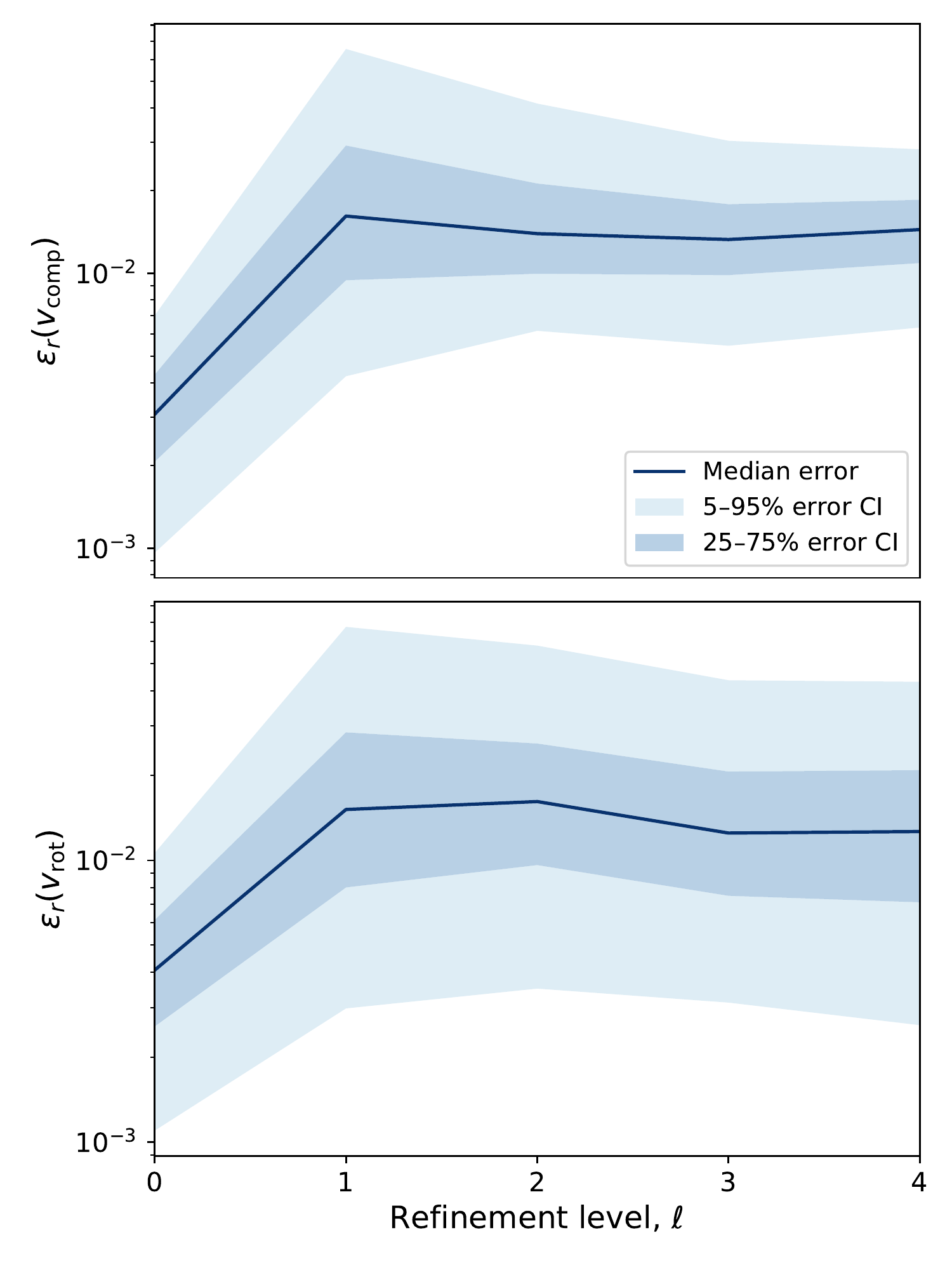}
\caption{Results from Test 4. Upper (\textit{lower}) panel presents the relative error in the reconstruction of the compressive (\textit{rotational}) velocity fields.}
\label{fig:test4_results}
\end{figure}

Last, for the exceedingly complex Test 4, since the actual decomposition is still known, we follow the same procedure as in Test 3. The detailed results, presented in the same way as for the previous test, can be found in Fig. \ref{fig:test4_results}. Compared to Tests $1-3$, in this case we find errors up to an order of magnitude higher, which is not surprising since in this case we have included a truly multi-scale velocity field, with signal spanning over almost 3 decades in spatial frequency. In any case, our code performs the HHD decomposition with median errors around $1\%$, and not exceeding $7\%$ even at 95-percentile level, despite the fact that we have introduced oscillations close to the grid nominal frequency. 

The results for this last test, together with the ones we present below in Sec. \ref{s:tests_masclet.performance}, show the ability of our algorithm to perform the HHD in challenging, very non-linear velocity fields.

\section{Application to a cosmological simulation}
\label{s:tests_masclet}

Last, this section focuses on the results of our HHD code when applied to the outcomes of a cosmological simulation, which is described in the paragraphs below. As opposed to the previous highly idealised tests $1-3$, the velocity field in a full-cosmological simulation exhibits a plethora of complex features due to the non-linear nature of the equations governing its evolution (see, e.g., \cite{Bertschinger_1998, Dolag_2008, Planelles_2015} for classical and recent reviews).

\subsection{Simulation details}
\label{s:tests_masclet.simulation}

The simulation analysed in this paper has been carried out with the cosmological code \textsc{MASCLET} \citep{Quilis_2004}, and has been already employed in a series of previous works \citep{Quilis_2017, Planelles_2018, Valles_2020}. Here we shall introduce the main details of the simulation, while some topics which are not intimately connected to the analyses in this paper can be found in more detail in the aforementioned references.

\textsc{MASCLET} is an Eulerian cosmological code, implementing \textit{high-resolution shock-capturing} techniques for the description of the gaseous component and an $N$-Body particle-mesh for dark matter (DM). Both components are built into an AMR scheme to gain spatial and temporal resolution in the regions of interest.

We have simulated a cubic domain of comoving side length $40 \, \mathrm{Mpc}$, assuming a flat $\Lambda$CDM cosmology with the following values of the cosmological parameters: $h \equiv H_0 /( 100 \, \mathrm{km\, s^{-1}\,Mpc^{-1}}) = 0.678$, $\Omega_m = 0.31$, $\Omega_b = 0.048$, $\Omega_\Lambda = 0.69$, $n_s = 0.96$ and $\sigma_8 = 0.82$, which are consistent with the latest values reported by the Planck mission \cite{Planck_2018_VI}. The domain is discretised in a base grid of $128^3$ cells, granting a harsh resolution of $\sim 310\, \mathrm{kpc}$ at the coarsest level. Regions with large gaseous and/or DM densities can get recursively refined following the AMR scheme. We allow $n_\ell = 9$ refinement levels, each one halving the cell side length with respect to the previous level, providing a peak resolution of $\sim 610\, \mathrm{pc}$. The peak DM mass resolution is $\sim 2 \times 10^6 M_\odot$, equivalent to filling the domain with $1024^3$ of such particles.

The simulation started at redshift $z=100$, with the initial conditions set up by a CDM transfer function \citep{Eisenstein_1998} and generated by a constrained realisation of the gaussian random field aimed to produce a massive cluster in the center of the computational domain by $z \sim 0$ \citep{Hoffman_1991}. By redshift $z \sim 0$, several massive clusters and groups have been formed. Besides gravity, the simulation accounts for a broad variety of feedback mechanisms, which are explained in greater detail in the cited previous works.

\subsection{Performance of the code}
\label{s:tests_masclet.performance}
\begin{figure}
\centering
\includegraphics[width=\linewidth]{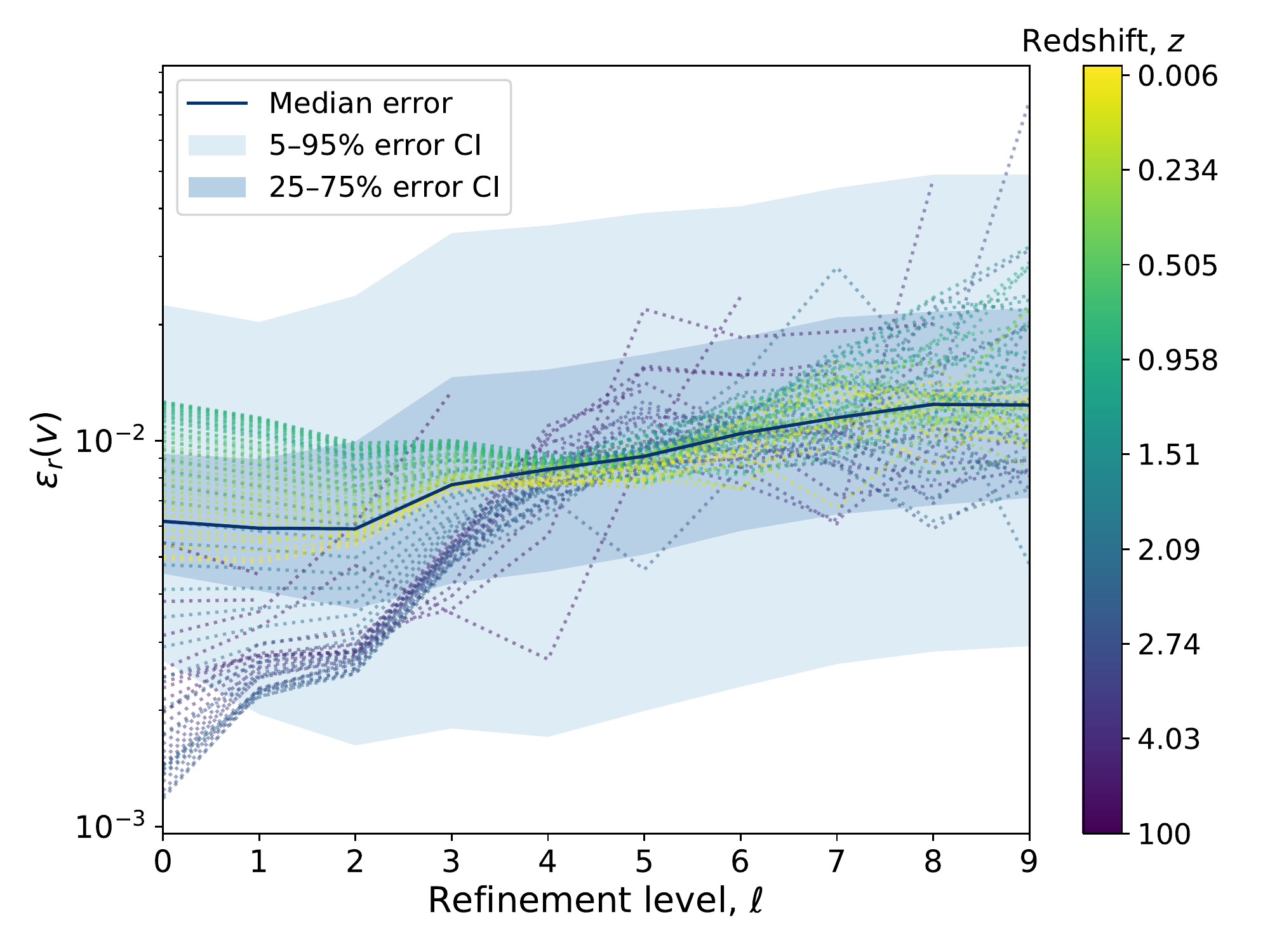}
\caption{Results from the tests of the HHD algorithm over \textsc{MASCLET} outputs. The blue solid line and the contours present the median over all the snapshots of the median relative error and the CIs defined as in the figures in Sec. \ref{s:tests}. The dotted lines represent the individual error statistics for each snapshot, with the colourscale encoding the redshift.}
\label{fig:test_masclet}
\end{figure}

We have run our HHD algorithm over 80 snapshots of the simulation described above, ranging from $z=100$ to $z=0$, and computed the cell-wise error\footnote{In this case, as we do not know beforehand the `true' decomposed velocity fields to compare with the reconstructed ones, we quantify the error by comparing the reconstructed total velocity field, $\vb{\tilde v} = \vb{\tilde v}_\mathrm{comp} + \vb{\tilde v}_\mathrm{rot}$ to the input one, as defined in Eq. \ref{eq:relative_error}. Since, by definition, $\vb{\tilde v}_\mathrm{comp}$ ($\vb{\tilde v}_\mathrm{rot}$) is the gradient of a scalar field (the rotational of a vector field), it is irrotational (solenoidal). We have, indeed, checked that our high-order derivatives verify this, typically much better than 1\textperthousand. Therefore, as the decomposition is unique, checking that $\vb{\tilde{v}} = \vb{v}$ proves the validity of the method.} as in Eq. (\ref{eq:relative_error}) and its corresponding percentiles (5, 25, 75 and 95), as done in the tests in Sec. \ref{s:tests}. Figure \ref{fig:test_masclet} presents the overall error statistics, defined as the median of the error statistics over all the code outputs. In order to keep track of the dispersion of the error statistics in different snapshots, we have also plotted the 80 individual error profiles in dotted lines, with the line colours encoding the redshift.

The median relative errors in describing the velocity field as the sum of a compressive and a rotational component are typically in the order of or slightly less than $1\%$ for all refinement levels, with only a small trend to increase the error with the refinement level. Even at the $95\%$ error percentile, the relative errors fall below $5\%$. At high redshift, the errors at low refinement levels ($\ell \leq 3$) tend to be smaller, most likely due to the fact that the velocity field does not present as complex features as it does at more recent redshifts due to its highly non-linear evolution. The behaviour at the most refined levels presents significant scatter and there is not a clear redshift evolution of the error, but its median magnitude keeps below a few percents in all snapshots. Thus, our algorithm is capable of providing a robust multi-scale decomposition of the velocity field in its compressive and rotational velocities, even on highly-complex, non idealised conditions.

\subsection{An example: velocity maps and profiles around a massive galaxy cluster}
\label{s:tests_masclet.maps}

\begin{figure*}
	\centering
	{\includegraphics[width=0.5\textwidth]{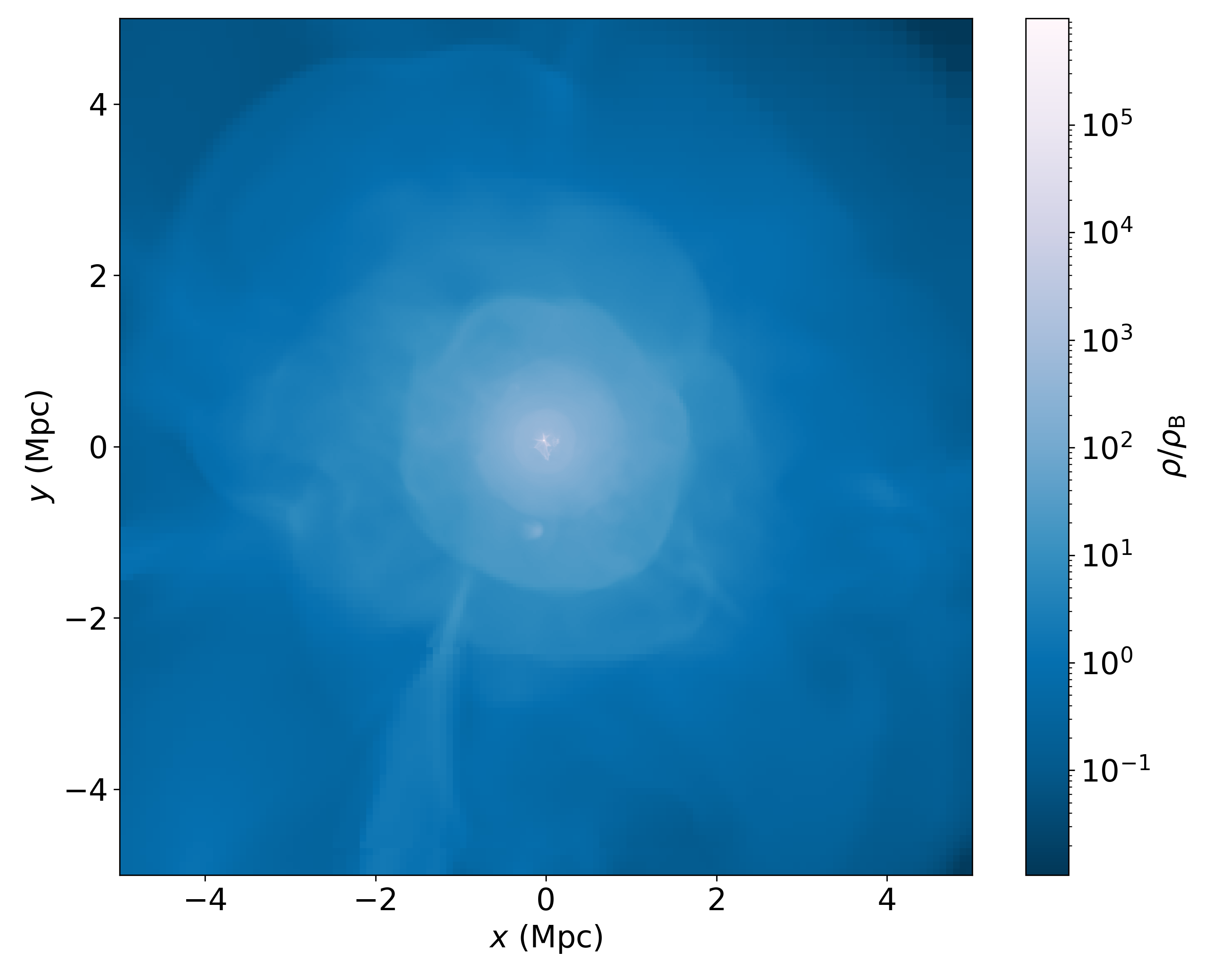}}~
	{\includegraphics[width=0.5\textwidth]{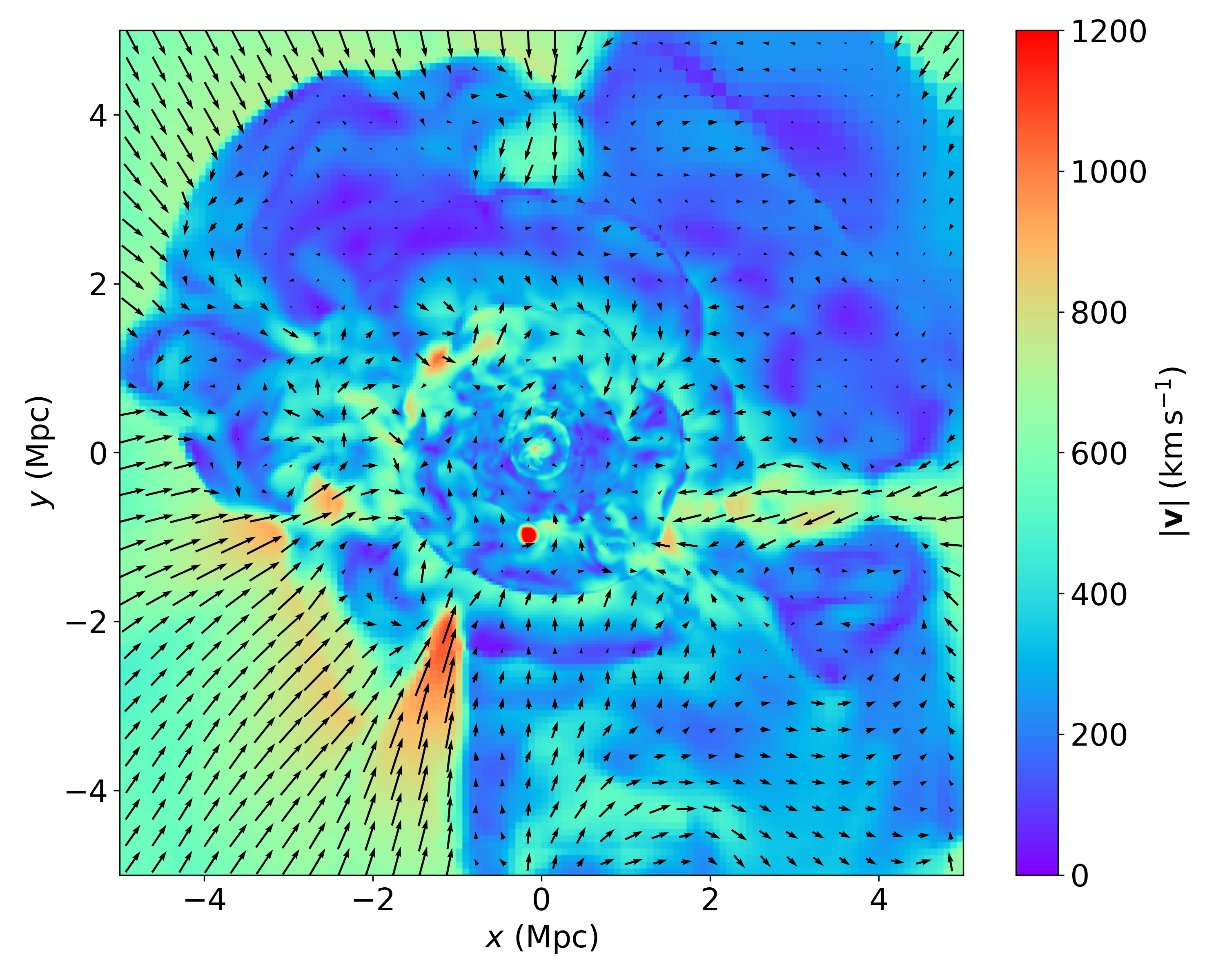}}
	{\includegraphics[width=0.5\textwidth]{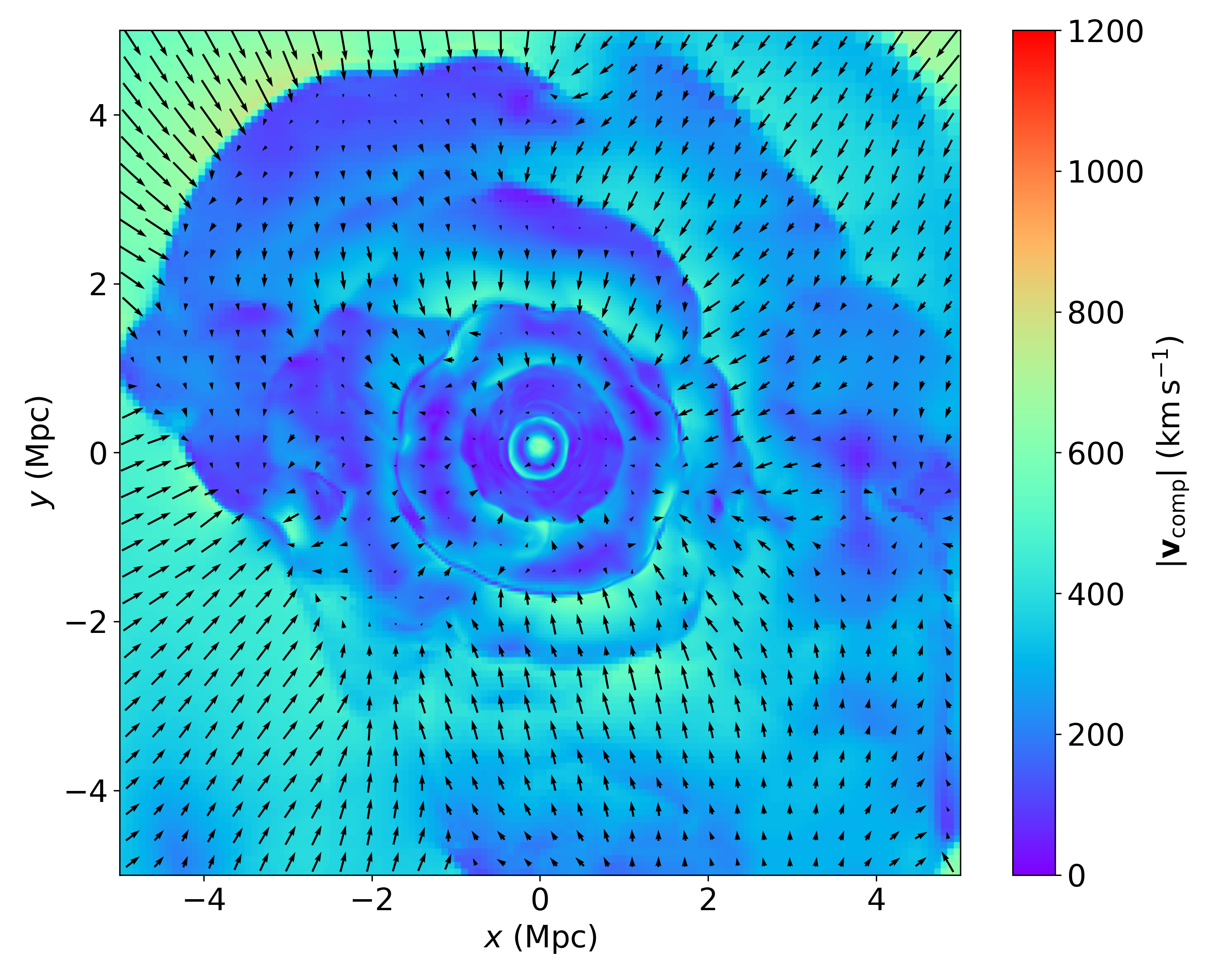}}~
	{\includegraphics[width=0.5
	\textwidth]{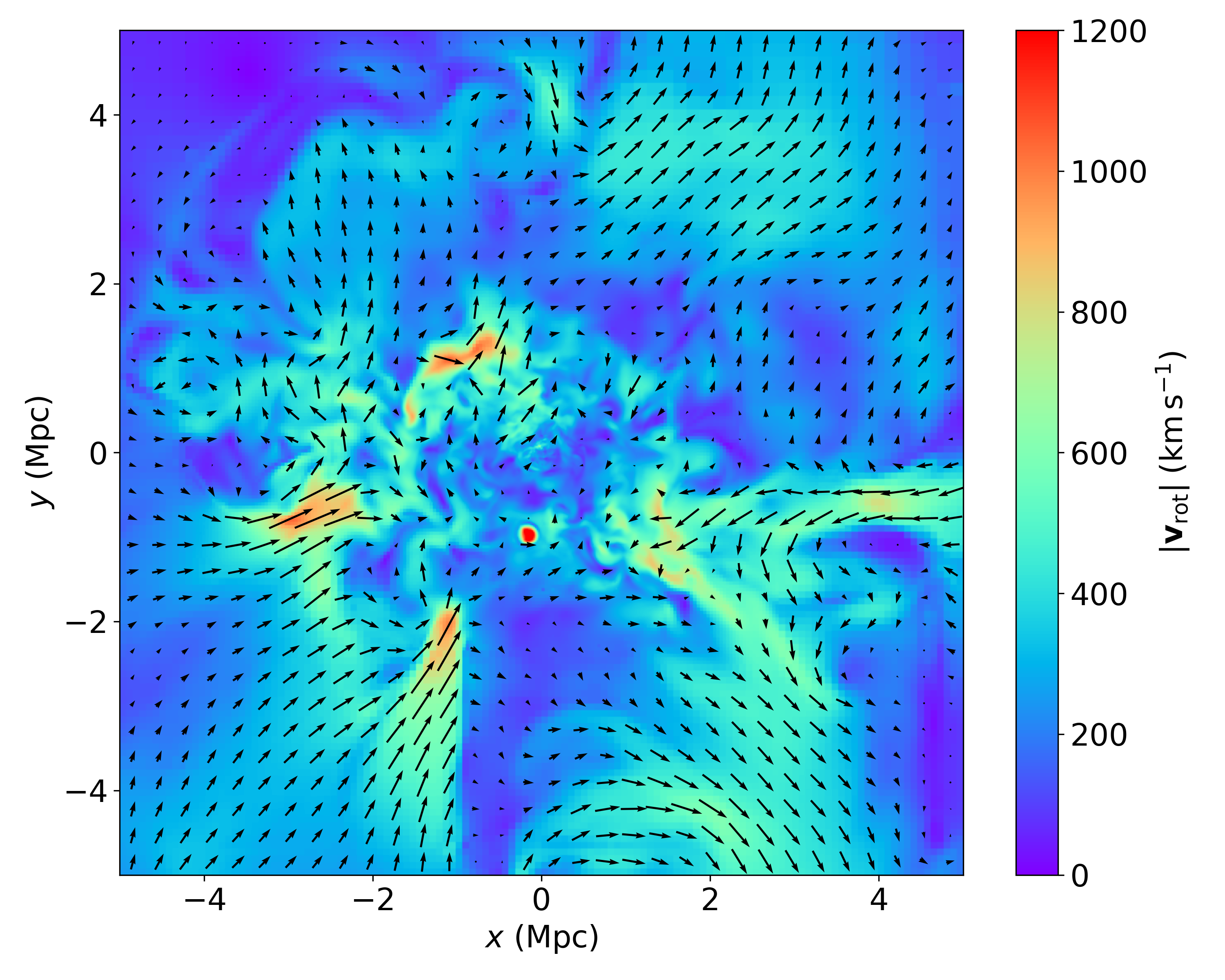}}
	\caption{Maps around a massive cluster. Each map is a mass-weighted projection of the same region around the cluster, 10 Mpc on each side and $\sim 150 \, \mathrm{kpc}$ thick, limiting the image resolution to $\sim 10 \, \mathrm{kpc}$. \textit{Top left:} Gas density (in units of the background density of the Universe). \textit{Top right:} Total velocity magnitude. \textit{Bottom left:} Compressive velocity magnitude. \textit{Bottom right:} Solenoidal velocity magnitude. The arrows in the velocity maps represent the projection of the corresponding velocity fields in the slice plane.}
	\label{fig:maps}
\end{figure*}

In order to exemplify the ability of the code to split the components of a highly complex velocity field, we focus on a massive galaxy cluster\footnote{This same object has been analysed in great detail in \cite{Planelles_2018} (focusing on its observational properties) and \cite{Valles_2020} (exploring its accretion history).}, with mass $M_\mathrm{vir} \simeq 4.83 \times 10^{14} M_\odot$ and radius $R_\mathrm{vir} \simeq 1.99 \, \mathrm{Mpc}$, at $z \simeq 0$. We present in Figure \ref{fig:maps} slices of gas density (top left panel), total velocity (top right), compressive velocity (bottom left) and solenoidal velocity (bottom right). The velocity maps show both magnitude (encoded in color) and direction in the slice plane (arrows).

The density map shows that, by $z \sim 0$, the cluster is, indeed, relatively relaxed, sitting in the gravitational potential dominated by the dark matter component, in an approximately spherical shape. Several density discontinuities can be easily discerned, mainly corresponding to (internal) merger shocks and (external) accretion shocks. A filament penetrating quite inner radii, of around $r \sim 1 \, \mathrm{Mpc}$, is also noticeable in the density plot. The total velocity field displays great complexity, especially in the inner regions of the cluster, where the variations occur on smaller scales (both because the dynamics are more complex and because, correspondingly, the resolution is higher). The filamentary structures appear to present high velocity magnitudes, mainly pointing radially, and velocity discontinuities, hinting the presence of shocks, are ubiquitous.

Compressive velocities show a nearly spherically symmetric pattern, as the cluster smoothly accretes gas from its surroundings. In the outskirts, baryonic matter is accelerated (and compressed) by the gravitational pull of the cluster, until it gets shocked causing the strong accretion shocks. In comparison, the post-shock medium presents very small compressive velocities, as the shock has effectively halted the accretion flows. Part of the energy corresponding to these compressive component gets thermalised, increasing the internal energy (and temperature) of the ICM. However, another important role of shocks is the generation of vorticity (see, e.g., \cite{Porter_2015}). Indeed, inside the external accretion shocks, the solenoidal component of the velocity fields gets relevant. Eddies develop on a wide range of scales, especially in the cluster central regions. It is also interesting to note how the infalling filament mentioned before presents high values of the rotational velocity, suggesting that matter is infalling following helicoidal trajectories. Being the main aim of this work is presenting the algorithm, we may defer a more in-depth analysis of these issues to future work.

\begin{figure}
	\centering
	{\includegraphics[width=\linewidth]{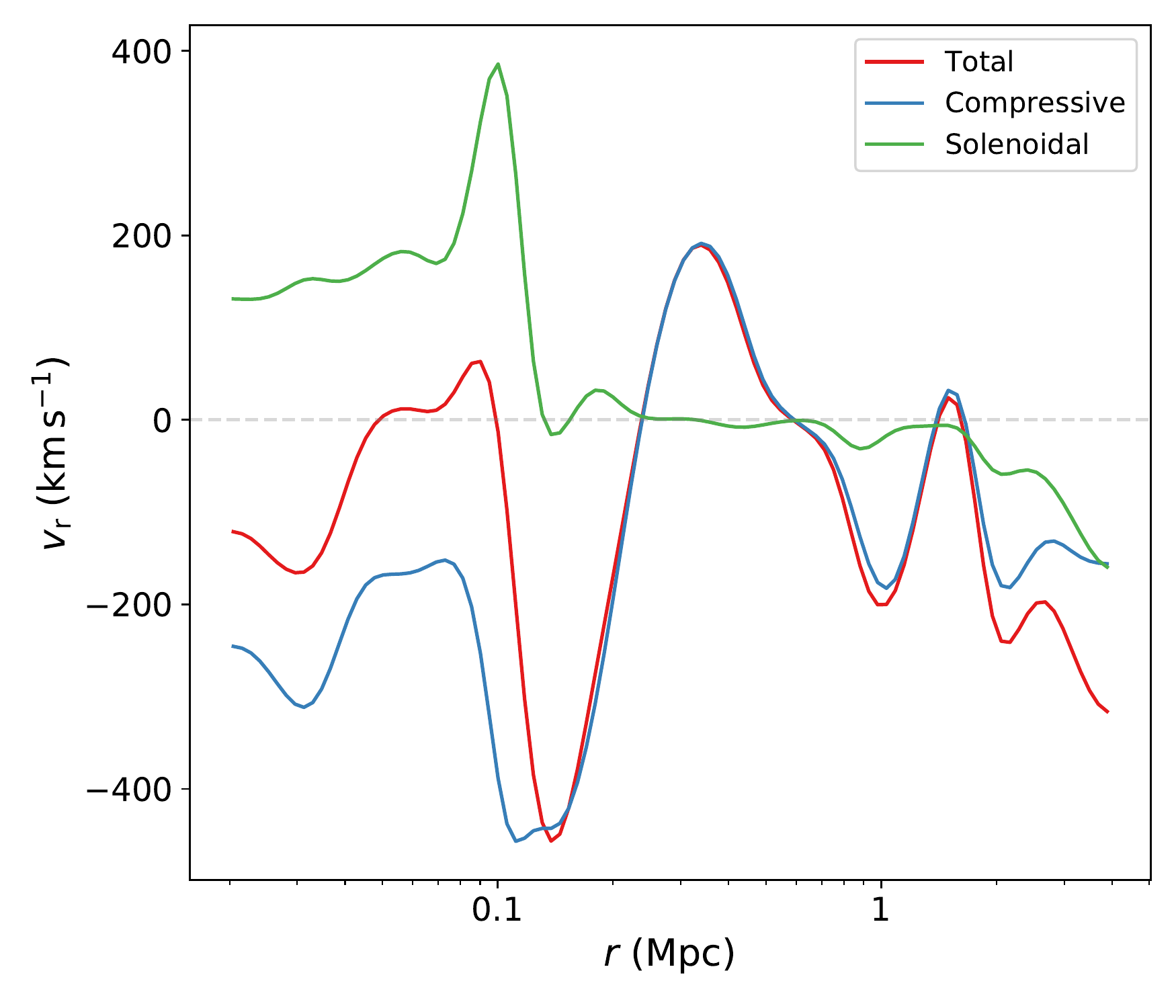}}
	\caption{Radial profiles of radial velocity for the total (\textit{red}), compressive (\textit{blue}) and solenoidal (\textit{green}) velocity components.}
	\label{fig:velocity_profiles}
\end{figure}

Complementarily, in Figure \ref{fig:velocity_profiles} we present the radial profile of radial total (red line), radial compressive (blue) and radial solenoidal (green) velocities, for $20 \, \mathrm{kpc} \lesssim r \lesssim{4 \, \mathrm{Mpc}} \simeq 2 R_\mathrm{vir}$. In the inner regions ($r\lesssim 0.1 \, \mathrm{Mpc}$) both strong radial compressive and solenoidal flows are present, bounded by an internal shock (clearly visible in the compressive velocity map of Fig. \ref{fig:maps}). For $r \gtrsim 0.1 \, \mathrm{Mpc}$, the radial compressive velocity clearly dominates. While solenoidal motions are still present, the fact that their radial component is close to zero suggests that these motions tend to occur along the tangential direction.

\section{Conclusions}
\label{s:conclusion}
In this paper, we have proposed a novel method to perform a Hemholtz-Hodge decomposition in AMR velocity fields, or virtually in any description which can be smoothed over an ad-hoc hierarchy of grids. Although our primary focus has been cosmological simulations of structure formation, the method is general and could be easily extended to any type of hydrodynamical simulation.

Previous works in the field of numerical cosmology typically use uniform grids and work straightforwardly in Fourier space. However, this procedure requires to perform constrained simulations (or resimulations) of specific objects of interest (e.g., a galaxy cluster) in order to be able to describe it with a uniform computational grid at a reasonable computational cost. Our algorithm, instead, can be applied to full-cosmological simulations, without the need of performing resimulations and keeping the full description at the maximum resolution at each position.

The performance of the code has been validated in a series of idealised tests, for which the analytical decomposition is known. Our algorithm has shown to succeed in disentangling the compressive and solenoidal velocity components and reconstructing the input velocity field, with typical errors in the order of 1\textperthousand$\,$ or below ($1\%$ in the more complex, ICM-like test). Our errors seem comparable to or even better than the ones displayed by the tests in \cite[Appendix A1.2]{Vazza_2017}.

For exceedingly complex velocity fields, like the ones generated by actual cosmological simulations at low redshifts, where turbulence is fully developed (e.g., \cite{Miniati_2014}) and velocity fluctuates on many different scales, our tests show that the decomposition can be brought about with median errors below $1\%$, even resolving scales smaller than the $\mathrm{kpc}$ in a domain of several tens of $\mathrm{Mpc}$ along each direction. 

This procedure, whose implementation has been made publicly available (see Sec. \ref{s:methods.numerics}), will allow us to further explore the dynamics of the turbulent velocity field in the ICM of large samples of clusters in future works.

\section*{Acknowledgements}
We gratefully acknowledge the anonymous referees for their valuable feedback, which has helped us to improve the quality of this manuscript. This work has been supported by the Spanish Ministerio de Ciencia e Innovaci\'on (MICINN, grants {AYA2016-77237-C3-3-P} and {PID2019-107427GB-C33}) and by the Generalitat Valenciana (grant PROMETEO/2019/071). Simulations have been carried out using the supercomputer Llu\'is Vives at the Servei d'Inform\`atica of the Universitat de Val\`encia. This research has made use of the following open-source packages: \textsc{NumPy} \citep{Numpy}, \textsc{SciPy} \citep{Scipy}, \textsc{matplotlib} \citep{Matplotlib} and \textsc{yt} \citep{yt}.

\bibliographystyle{elsarticle-num} 
\bibliography{cpc_vortex.bib}

\end{document}